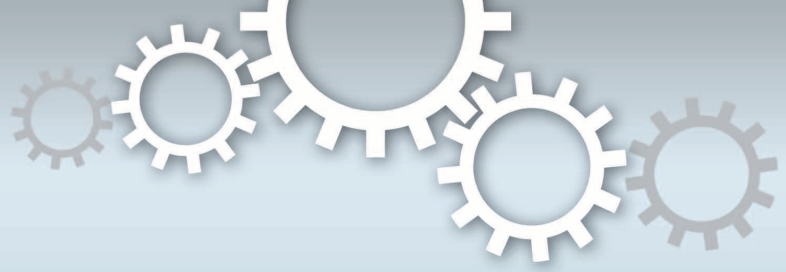

# SCIENTIFIC REPORTS

# High-efficient thermoelectric materials: The case of orthorhombic IV-VI compounds

Guangqian Ding, Guoying Gao & Kailun Yao

School of Physics & Wuhan National High Magnetic Field Center, Huazhong University of Science and Technology, Wuhan 430074, China.

Improving the thermoelectric efficiency is one of the greatest challenges in materials science. The recent discovery of excellent thermoelectric performance in simple orthorhombic SnSe crystal offers new promise in this prospect [Zhao *et al.* Nature 508, 373 (2014)]. By calculating the thermoelectric properties of orthorhombic IV-VI compounds GeS,GeSe,SnS, and SnSe based on the first-principles combined with the Boltzmann transport theory, we show that the Seebeck coefficient, electrical conductivity, and thermal conductivity of orthorhombic SnSe are in agreement with the recent experiment. Importantly, GeS, GeSe, and SnS exhibit comparative thermoelectric performance compared to SnSe. Especially, the Seebeck coefficients of GeS, GeSe, and SnS are even larger than that of SnSe under the studied carrier concentration and temperature region. We also use the Cahill's model to estimate the lattice thermal conductivities at the room temperature. The large Seebeck coefficients, high power factors, and low thermal conductivities make these four orthorhombic IV-VI compounds promising candidates for high-efficient thermoelectric materials.

T hermoelectric materials which can directly convert waste heat into electrical energy have been a hot spot in research community especially under the challenges of energy crisis and global warming. The efficiency of thermoelectric materials is given by the dimensionless figure of merit[1]:

$$ZT = \frac{S^2\sigma}{\kappa_e + \kappa_l}T. \qquad (1)$$

Where $S$, $\sigma$, $\kappa_e$, $\kappa_l$, and $T$ are the Seebeck coefficient, electrical conductivity, electrical thermal conductivity, lattice thermal conductivity, and temperature, respectively. A large value of $ZT$ means a higher thermoelectric conversion efficiency which requires a high power factor ($S^2\sigma$) along with a low thermal conductivity ($\kappa_e + \kappa_l$). Unfortunately, the thermoelectric figure of merit of conventional three-dimensional crystalline systems can not be infinitely increased because these parameters are restricted: a large Seebeck coefficient needs low carrier concentration, which results in low electrical conductivity; a large electrical conductivity always comes with a high electrical thermal conductivity. Most works in the past decade focused on improving $ZT$ by several approaches[2]. For example, enhancing the power factor by band engineering (degeneration of multiple valleys, electronic resonance states, and depressing bipolar effect, etc.)[3–5] and reducing lattice thermal conductivity by all length-scale hierarchical structuring (nanoscale precipitates and superlattice)[6,7]. Although these approaches may greatly enhance $ZT$ in theory or experiment, it is difficult to realize them in practical applications due to the complicated process and high cost. Therefore, searching for new three-dimensional materials is still an important route to obtain high-performance thermoelectricity.

Very recently, ultralow thermal conductivity and high thermoelectric figure of merit have been experimentally found in three-dimensional bulk SnSe by Zhao *et al.*[8]. An unprecedented $ZT$ of 2.6 at 973 K was found in SnSe single crystal along the $b$ axis owing to it's exceptionally low thermal conductivity. Afterwards, Sassi *et al.*[9] and Chen *et al.*[10] also reported high $ZT$ values and low thermal conductivities in polycrystalline SnSe, but the $ZT$ values are lower than those measured in single crystal SnSe by Zhao *et al.*[8], and there is also large discrepancy between the thermal conductivity measured on single crystals[8] and polycrystals[9,10]. These works gave us inspiration that high thermoelectric performance can be realized in simple bulk materials without complicated crystal structure and heavy atoms instead of complex strategies to nanostructuring[1]. Moreover, further studies on the thermal conductivity of SnSe are necessary in order to understand the different measured results of thermal conductivities between crystals[8] and polycrystals[9,10]. SnSe undergoes a reversible phase transition from low-temperature phase (*Pnma* space group) to high-temperature phase (*Cmcm* space group) with transition temperature around 800 K. The *Pnma* phase contains eight atoms and two adjacent double layers in the primitive cell. One atom in a single layer is collected with three nearest neighbors, which forms zigzag chains[11]. Since the low thermal conductivity and high *ZT* have been found in SnSe, this kind of special layered and anisotropic zigzag structure is expected to exhibit novel thermoelectric properties. Thus, it is necessary and interesting to explore the thermoelectric performance of new materials with the same or similar crystal structure of SnSe.

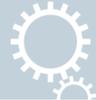

Experimentally, the IV-VI compounds GeS, GeSe, and SnS all crystallize in GeS-type structure which belongs to the orthorhombic structure with space group *Pnma*[11]. However, studies on the thermoelectric properties of these orthorhombic compounds except for SnSe are few[11–13], and there is not a comparison of thermoelectric performance among them. Recently, Tan *et al.* experimentally found a modest *ZT* of 0.6 at 873 K in polycrystalline SnS[14], which is comparable to those measured in polycrystalline SnSe[9,10]. Here, in order to search for more high-performance thermoelectric materials, we investigate and compare the thermoelectric properties of the orthorhombic IV–VI compounds by using the density function theory combined with the Boltzmann transport equation. We reveal that GeS, GeSe, and SnS show similar thermoelectric properties to that of SnSe: large Seebeck coefficient, high power factor, and low thermal conductivity. Importantly, the Seebeck coefficients of GeS,GeSe, and SnS are even larger than that of SnSe under the studied carrier concentration and temperature region. Moreover, the Seebeck coefficients and power factors can be greatly enhanced by both n-type and p-type doping.

**Computational details.** Our calculations are based on the density function theory combined with the Boltzmann transport equation. Firstly, structure optimization and electron energy calculations are performed by the full-potential linearized augmented plane-wave method (WIEN2K code)[15], which is a powerful tool to deal with the electronic structure of various solid systems[16,17]. In this method, the exchange-correlation functional is adopted with the generalized gradient approximation in the form of Perdew-Bueke-Ernzerhof. The plane-wave cutoff is set as $R_{mt}K_{max} = 9$ and the total energy is converged to within $10^{-5}$ Ry. Then, based on the calculated electronic energy, we obtain the thermoelectric transport coefficients by solving the linearized Boltzmann equation in relaxation time approximation within Boltztrap code[18]. In order to obtain reliable results, the *k* points in the irreducible Brillouin zone are tested to use at least 60000 points. Boltzmann transport calculations have long been used for bulk semiconductors and show reliable results[19,20]. The phonon dispersion relations are obtained from the phonopy code[21], which realizes the Parlinski-Li-Kawazoe method, conducted in supercell approach with the finite displacement method[22]. In order to obtain reliable phonon spectrum, a $2 \times 4 \times 4$ supercell is employed for force constant calculations and at least 1000 *k* points are used. Here, we evaluate the room temperature minimum lattice thermal conductivities of these compounds by the Cahill's[23] method, which borrowed from Debye model and has been successfully used in predicting the lattice thermal conductivity[24,25]. The crystal structures of GeS, GeSe, SnS, and SnSe have been studied detailed in previous works[11,26]. In this work, we will mainly concern the thermoelectric properties in the moderate temperature region (*Pnma* phase) because the measured temperatures of phase transition in these compounds are around or above 800 K[26].

## Results and discussion

The optimized lattice constants and calculated band gaps are shown in Table 1, which are in good agreement with the experimental values[11]. Obviously, GeS shows the largest gap of 1.25 eV while SnSe has the lowest gap of 0.69 eV, and a little gap difference between GeSe and SnS is found. These gaps are much higher than those of traditional thermoelectric materials (0.105 eV for $Bi_2Te_3$ and 0.14 eV for $Sb_2Te_3$)[7], which would make great difference in thermoelectric properties. A large gap may cause lower carrier concentration around the Fermi level. As transport coefficients mainly depends on the electronic properties. It would result in large Seebeck coefficients, and also, large power factors can be obtained by reasonable doping.

We firstly calculate the thermoelectric properties of orthorhombic SnSe in order to compare our calculated results with the recent experiment[8]. The electrical conductivities and Seebeck coefficients of SnSe along different crystallographic directions as a function of temperature are shown in Figs. 1(a) and 1(b). Temperatures from 300 K to 750 K are considered in our calculations, which are below the phase transition temperature around 800 K[26]. We have assumed a series of carrier concentrations ranging from $1 \times 10^{17}/cm^3$ to $1 \times 10^{18}/cm^3$ according to the experimental Hall measurements[8]. We finally adopt the value of $6 \times 10^{17}/cm^3$ because the obtained thermoelectric properties with it are better consistent with experiment results. As shown in Fig. 1(b), the calculated Seebeck coefficients along b and c axis are in good agreement with experimental data[8], while the behavior of a axis is lower than the experimental value and decreases rapidly above 450 K and finally down to negative value when the temperature exceeds 600 K. We attribute this rapid decrease and sign modification along the a axis to the increase of thermally excited negative electron. We note that the recent experimental Seebeck coefficients on polycrystalline SnSe showed comparable values[9], however, unlike the result of Zhao *et al.*[8], the Seebeck coefficient doesn't show constant but further decreases above the transition temperature[9], which remains to be understood. Fig. 1(a) shows the anisotropic electrical conductivities in the temperature range considered. Under the relaxation-time approximation in Boltzmann transport calculations, the Seebeck coefficient(*S*) is independent of the relaxation time $\tau$, whereas the electrical conductivity ($\sigma$) depend linearly on $\tau$. Furthermore, the electrical thermal conductivity ($\kappa_e$) is obtained by the Wiedemann-Franz equation ($\kappa_e = L\sigma T$, L: Lorentz factor), thus, $\kappa_e$ is also relevant to $\tau$. Usually, it is difficult to determine relaxation time in a bulk material because of the complex scattering mechanism (phonon scattering, carrier scattering, defect scattering, and boundary scattering). Here, we fix the relaxation time at $1 \times 10^{-14}$ s in our calculations by comparing the calculated results with experiment. Yabuuchi *et al.*[27] also adopt this value in their calculations for its rationality. Electrical conductivities show weakly decrease in the 300–500 K range but increase rapidly for more carriers are excited as temperature shifts to higher value, which corresponds to the slowly increase and quickly decrease of

**Table 1 | Optimized lattice constants and calculated band gaps of GeS, GeSe, SnS, and SnSe**

| Composition | a(Å) | b(Å) | c(Å) | gap(eV) |
|---|---|---|---|---|
| GeS | 10.635 | 3.668 | 4.738 | 1.25 |
| GeSe | 10.954 | 3.828 | 4.492 | 0.87 |
| SnS | 11.392 | 4.043 | 4.353 | 0.91 |
| SnSe | 11.754 | 4.203 | 4.441 | 0.69 |

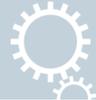

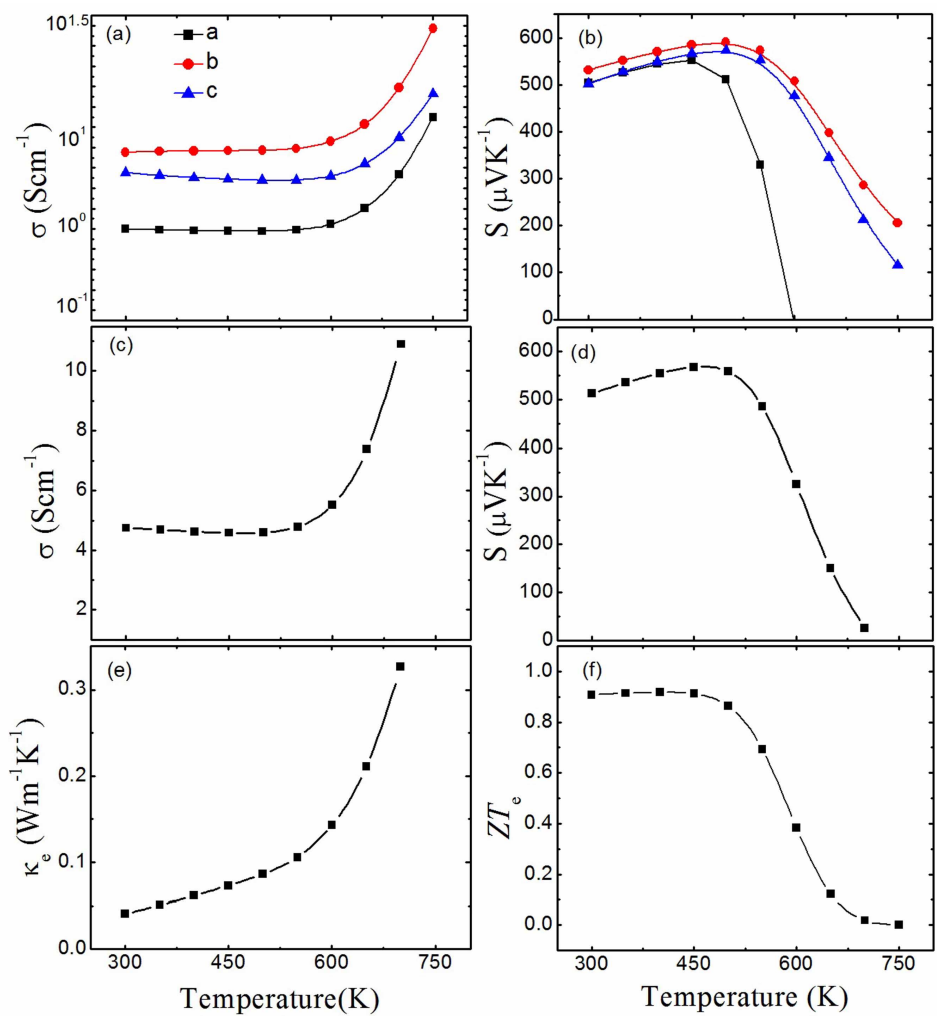

**Figure 1 | Thermoelectric parameters as a function of temperature for SnSe:** (a) Anisotropic electrical conductivity, (b) Anisotropic Seebeck coefficient, (c) Electrical conductivity, (d) Seebeck coefficient, (e) Electrical thermal conductivity, and (f) Thermoelectric figure of merit (only electrical thermal conductivity concerned).

Seebeck coefficients in the temperature region. Figs. 1(c) and 1(d) give the total electrical conductivities and Seebeck coefficients, which equal to the average values of three crystal directions. The largest Seebeck coefficient of 570 $\mu V/K$ is found at 450 K, and electrical conductivity has an almost constant value near 5 Scm$^{-1}$ in the room temperature region. Fig. 1(e) presents the low electrical thermal conductivity of SnSe, which ranges from 0.04 to 0.3 Wm$^{-1}$K$^{-1}$ as temperature increase.

When towards the thermoelectric figure of merit, the relaxation time $\tau$ will also be an undetermined parameter because of the relaxation-time approximation. Here, we rewrite the equation (1) as,

$$ZT = \frac{S^2\sigma T}{\kappa_e}\frac{\kappa_e}{\kappa_e + \kappa_l}. \qquad (2)$$

Thus, the ratio $(ZT_e = S^2\sigma T/\kappa_e)$ is independent of the relaxation time $\tau$ and is an upper limit to the thermoelectric figure of merit, which only ignores the thermal conductivity from lattice contribution. The value of $ZT_e$ approaches $ZT$ if the lattice contribution to the thermal conductivity is insignificant compared with the electronic term. In very low temperature, a very few electrons are excited, which leads to low electrical thermal conductivity. Thus, the thermal conductivity may dominated by lattice term and estimating $ZT$ needs to better consider the lattice thermal conductivity. As temperature increases especially when higher than room temperature, a large number of excited electrons will lead to the increase of electrical thermal conductivity, while the lattice contribution is decreased because of the increasing ratio of phonon scattering resulting from severe lattice vibration. Therefore, the ratio $ZT_e$ is a good estimate of $ZT$ as temperature increase to high value. As shown in Fig. 1(f), a $ZT_e$ platform about 0.92 at 300 K to 450 K is found, and it rapidly decrease down to near zero in higher temperature because of the rapidly increasing electrical thermal conductivity. It is evident that the *Pnma* phase of SnSe may reach its upper limit $ZT$ in the moderate temperature.

We adopt the same carrier concentration of $6 \times 10^{17}/cm^3$ to calculate the transport coefficients of GeS, GeSe, and SnS for better comparison with SnSe. Remarkably, the Seebeck coefficients of GeS, GeSe, and SnS (Fig. 2(b)) are larger than that of SnSe (Fig. 1(d)) in the whole temperature range. In Fig. 2(b), GeS shows the largest Seebeck coefficient while SnS has the lowest one in the 300–600 K range. However, the Seebeck coefficient of GeSe decreases faster compared to that of SnS when the temperature exceeds 600 K. The maximum values of Seebeck coefficients for GeSe, SnS, and GeS (706 $\mu V/K$ at 560 K for GeSe, 668 $\mu V/K$ at 650 K for SnS, and 769 $\mu V/K$ at 760 K for GeS, respectively) shift to higher temperature compared with SnSe (570 $\mu V/K$ at 450 K). In Fig. 2(c), electrical thermal conductivities of GeS, GeSe, and SnS are independent of temperature and much lower than that of SnSe (see Fig. 1(e)) in 300–750 K range, and the values keep nearly under 0.1 Wm$^{-1}$K$^{-1}$ until 750 K, which indicate more excellent thermal stability of GeS, GeSe, and SnS than SnSe. It can be readily seen that rapid increase of electrical thermal conductivities occurs above 750 K. This behavior corresponds to the increase of electrical conductivities shown in Fig. 2(a) due to the proportional relationship between electrical conductivity and electrical thermal conductivity. It is shown in Fig. 1(d) that the trends of $ZT_e$ of GeS, GeSe, and SnS are similar to that of SnSe. However, there are two differences we make clear: () the temperature regions of the $ZT_e$ platforms are wider than SnSe especially for GeS (from 300 K to 650 K); () it can be seen that the $ZT_e$ platforms of GeS, GeSe, and SnS are all higher than that of SnSe, and the highest value approaching to 1 is found in SnS. Comparatively speaking, GeS, GeSe, and SnS may show better thermoelectric performance than SnSe.

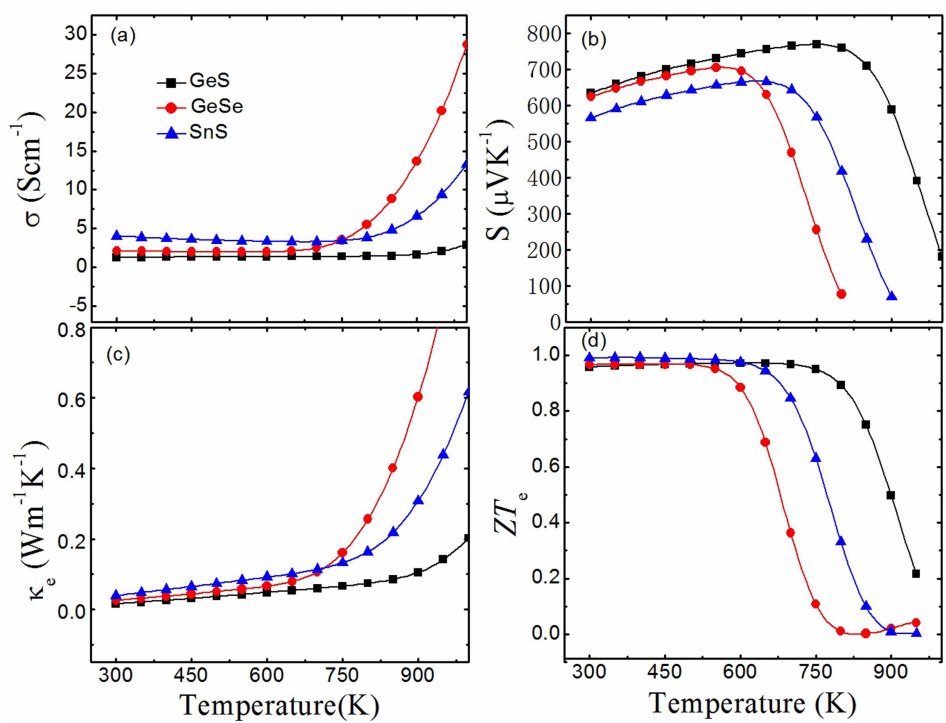

**Figure 2 | Thermoelectric parameters as a function of temperature for GeS, GeSe, and SnS:** (a) Electrical conductivity, (b) Seebeck coefficient, (c) Electrical thermal conductivity, and (d) Thermoelectric figure of merit (only electrical thermal conductivity concerned).

In Fig. 3, the calculated Seebeck coefficients, power factors, and $ZT_e$ under different temperatures (300 K and 600 K) as a function of chemical potential are presented in order to optimize the thermoelectric performance of these compounds. Within the rigid-band picture, the chemical potential indicates the doping level of the compound. As for n-type doping, the Fermi level shifts up, which corresponds to positive $\mu$. Instead, the Fermi level shifts down for p-type doping and the corresponding $\mu$ is negative[28]. Surprisingly, $S$ is greatly enhanced in a narrow region around $\mu = 0$, which indicates that a quite remarkable value of $S$ can be achieved through small n-type or p-type doping. It is evident that the maximum $|S|$ move towards the gap edges from SnSe to GeSe, SnS, and GeS as their band gaps increase. At 300 K, GeS almost reaches the largest value of 2000 $\mu V/K$ while SnSe shows lowest value of 930 $\mu V/K$, GeSe and SnS show almost the same maximum $|S|$ because of their small difference in gaps (see Table 1). However, these values almost halved at 600 K. It is worth noting that $|S|$ at 300 K comes closer to $S = 0$ than that at 600 K at higher $|\mu|$, which shows that higher power factor at 600 K in these chemical potential regions would be found. Obviously, large $|S|$ leads to large power factor for both n-type and p-type doping, and n-type doping performs better results than p-type doping. On the contrary, the peaks of power factors under 600 K are much higher than those under 300 K due to the higher $|S|$ in higher chemical potential regions. These optimized power factors for GeS, GeSe, and SnS are much larger than that of SnSe which has been obtained by recent experiment[8]. We summarize in Table 2 the peaks of power factors and corresponding carrier concentrations for both n-type and p-type doping at 300 K and 600 K of these compounds. The carrier concentrations are calculated from the Hall coefficients, based on the following formula[29],

$$N = \frac{1}{R_H e}. \qquad (3)$$

Here, $N$ and $R_H$ are the carrier concentration and Hall coefficient, respectively. These carrier concentrations are in the reasonable doping level[30], and it is desirable to dope these compounds in the range of $10^{20}$–$10^{21}$ cm$^{-3}$ to achieve high power factors. However, high power factor doesn't mean corresponding high thermoelectric figure of merit because of the high electrical thermal conductivity. As shown in Fig. 2, the $ZT_e$ of these compounds reach their upper limit in smaller $|\mu|$ compared with power factors. Higher $ZT_e$ at 300 K for these compounds are found, and with the highest value of 1.15 in GeSe, which further indicate the better performance in moderate temperature for the *Pnma* phase of these IV–VI compounds.

Finally, in order to give a theoretical prediction of the lattice thermal conductivity, we calculate the phonon dispersion relations of GeS, GeSe, SnS, and SnSe along several high symmetry lines as shown in Fig. 4 ($\Gamma$–X Brillouin zone direction represents the a crystal axis, $\Gamma$–Y the b axis, and $\Gamma$–Z the z axis, respectively). There is little difference in these dispersion relations except for the maximum frequency of vibrational modes. Since the primitive cells of these compounds contain eight atoms, twenty four vibrational modes can be found, including three acoustic branches and twenty one optical branches. The acoustic branches describe the vibration of the center of mass, and the optical branches describe the relative vibration between two atoms. In most cases, the frequency of the optical modes approximate to constant within the range of wave vector, leading to low group velocity in these phonons. Thus, the optical modes will not be interested in some cases because their contribution to thermal conductivity is limited. Considering Cahill, whose theoretical model also ignores the contribution of optical branches, and is thus an approximate value approaching to the minimum lattice thermal conductivity[23]. However, unlike half-Heusler alloys[31], these compounds contain a large number of optical modes, which may cause an obvious underestimation of lattice thermal conductivity in Cahill's model. Even so, we can also obtain a predictive result. Here, we use Cahill's model to theoretically predict the minimum lattice thermal conductivities at room temperature 300 K. The formula of Cahill's model is given by Ref. 23,

$$\kappa_{\min} = (\frac{\pi}{6})^{1/3} k_B n^{2/3} \sum_{i=1}^{3} v_i (\frac{T}{\theta_i})^2 \int_0^{\theta_i/T} \frac{x^3 e^x}{(e^x-1)^2} dx. \quad (3)$$

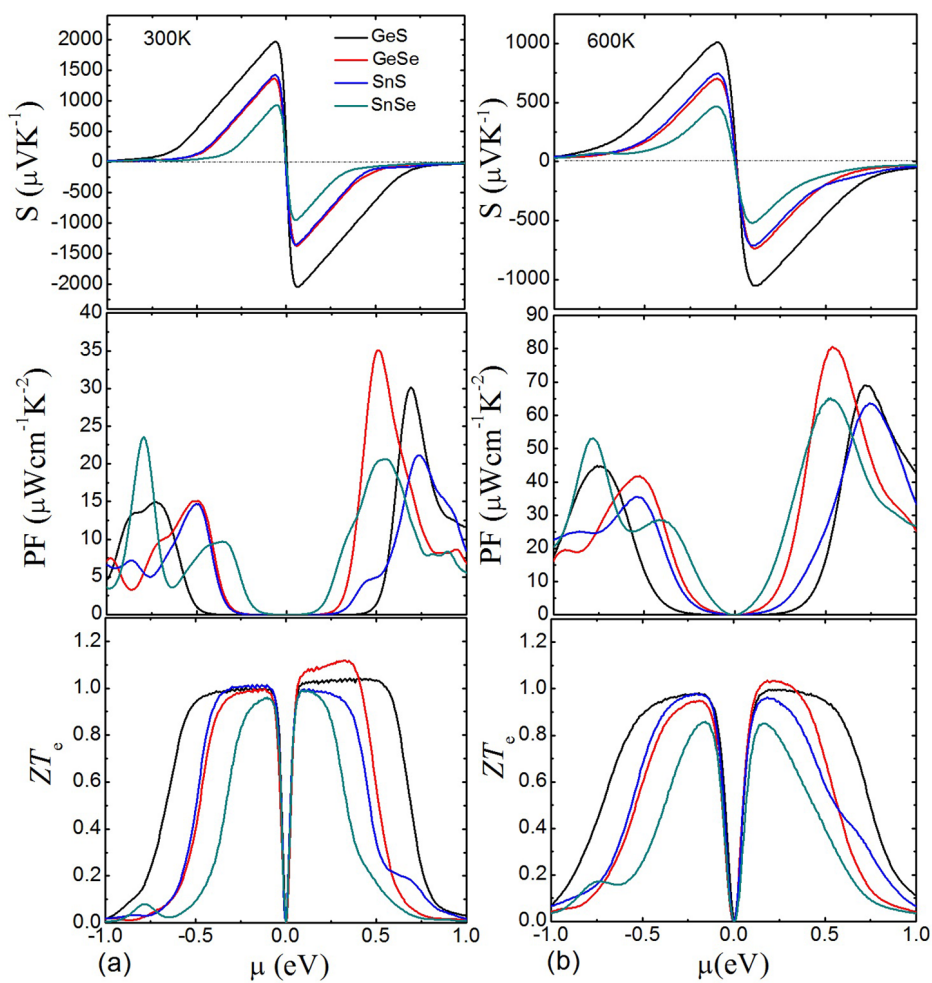

**Figure 3 | Seebeck coefficients, power factors (PF), and thermoelectric figure of merit (only electrical thermal conductivity concerned) as a function of chemical potential for GeS, GeSe, SnS, and SnSe under different temperatures (left part for 300 K and right part for 600 K).**

Where $k_B$ is the Boltzmann constant, and $n$ is the number density of atoms. The sum is taken over the three acoustic modes (two transverse and one longitudinal). $v_i$ is the phonon velocity, and $\theta_i$ is the debye temperature, $\theta_i = v_i(\hbar/k_B)(6\pi^2 n)^{1/3}$. The phonon velocities are the slope of the acoustic phonon dispersions around the $\Gamma$ point, which is given by,

$$v_i = \frac{\partial \omega_i(k)}{\partial(k)}. \quad (4)$$

Where $i$ represents $LA$, $TA$, and $TA'$. The calculated results are given in Table 3. The phonon velocities for each acoustic mode are the average of three crystal directions. As for SnSe, the calculated room temperature $\kappa_{min}$ is 0.32 Wm$^{-1}$K$^{-1}$, which is underestimated compared with Zhao's experimental measurement[8] and Carrete's calculation[32] due to the ignorance of optical modes contributions in the Cahill's model. What's more, we find the detailed study on the lattice thermal conductivity presented by Carrete et al.[32] somewhat contradicts to Zhao's measurement[8]. Moreover, both Sassi et al.[9] and Chen et al.[10] shown that the thermal conductivity measured by Zhao et al.[8] is underestimated and doubt the achieved high *ZT*. These differences need further theoretical and experimental investigations. The $\kappa_{min}$ of GeS, GeSe, and SnS are still much low but all higher than that of SnSe. While the $\kappa_{min}$ is underestimated, these compounds are also believed to show lower lattice thermal conductivity than other thermoelectric materials such as Bi$_2$Te$_3$ and TiNiSn[33,34].

**Table 2 | The peaks of power factors (PF) and corresponding carrier concentrations (N) per cubic centimeter for both n-type and p-type doping of GeS, GeSe, SnS, and SnSe at 300 K and 600 K**

| | 300 K | | | | 600 K | | | |
|---|---|---|---|---|---|---|---|---|
| | n-type | | p-type | | n-type | | p-type | |
| Composition | PF | N | PF | N | PF | N | PF | N |
| GeS | 30.25 | $-6.8 \times 10^{20}$ | 15.03 | $7.3 \times 10^{20}$ | 69.35 | $-4.0 \times 10^{20}$ | 44.91 | $1.5 \times 10^{21}$ |
| GeSe | 35.03 | $-1.2 \times 10^{20}$ | 15.02 | $1.4 \times 10^{20}$ | 80.47 | $-1.4 \times 10^{20}$ | 41.94 | $8.0 \times 10^{20}$ |
| SnS | 21.25 | $-7.0 \times 10^{20}$ | 14.70 | $1.3 \times 10^{20}$ | 63.41 | $-9.0 \times 10^{20}$ | 36.01 | $3.0 \times 10^{20}$ |
| SnSe | 20.07 | $-6.7 \times 10^{20}$ | 23.48 | $6.8 \times 10^{21}$ | 64.64 | $-5.8 \times 10^{20}$ | 52.52 | $7.1 \times 10^{21}$ |

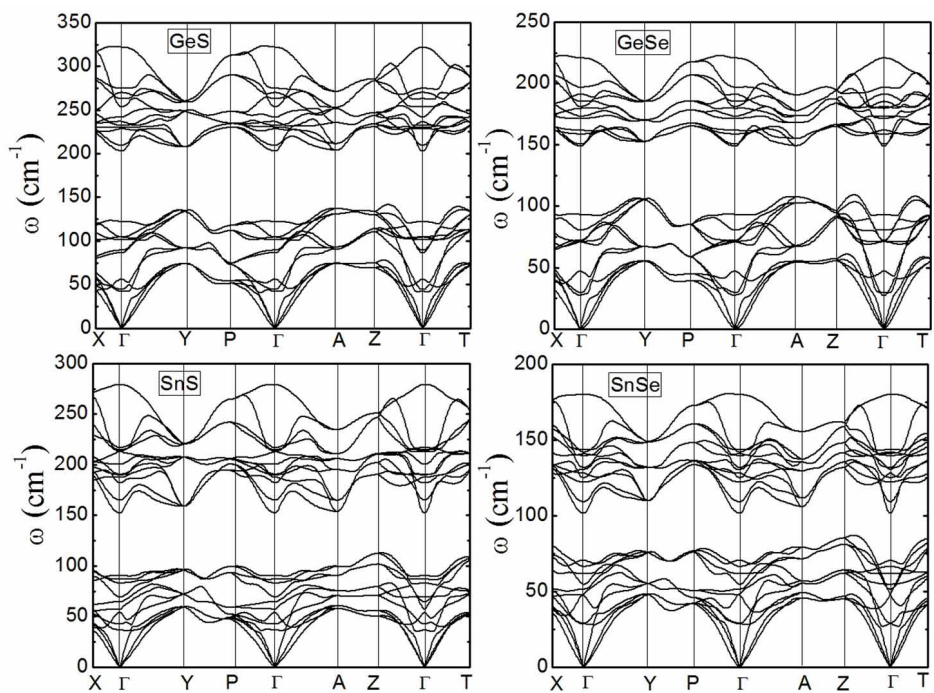

**Figure 4 | Phonon dispersion relations for GeS, GeSe, SnS and SnSe with *Pnma* phase along several high symmetry lines.**

**Table 3 | Calculated phonon velocities (two transverse and one longitudinal) and theoretical minimum lattice thermal conductivities at 300 K of GeS, GeSe, SnS, and SnSe**

|  | $n$ ($10^{22}$ cm$^{-3}$) | $v_{TA}$ (ms$^{-1}$) | $v_{TA}$ (ms$^{-1}$) | $v_{LA}$ (ms$^{-1}$) | $\kappa_{min}$ (Wm$^{-1}$K$^{-1}$) |
|---|---|---|---|---|---|
| GeS | 4.6 | 1647 | 2447 | 3412 | 0.52 |
| GeSe | 4.1 | 1138 | 2124 | 2820 | 0.39 |
| SnS | 3.9 | 1537 | 2343 | 3368 | 0.45 |
| SnSe | 3.6 | 1121 | 1568 | 2412 | 0.32 |

## Conclusion

In summary, motivated by the recent experiment that the unprecedented ultralow thermal conductivity and high thermoelectric figure of merit in SnSe single crystal, we have used the first-principles combined with the Boltzmann transport theory to investigate the thermoelectric properties of four orthorhombic IV–V compounds. It is found that GeS, GeSe, and SnS show comparative thermoelectric properties compared with SnSe. Large Seebeck coefficients, high power factors, and low thermal conductivities can be achieved in these compounds, which make them promising candidates for high efficient thermoelectric materials. The present work provides a detailed understanding of the excellent thermoelectric properties of these IV–VI compounds, and will stimulate experimental and theoretical studies of high-efficient thermoelectric materials.

## Acknowledgments

This work was supported by the National Natural Science Foundation of China under Grant Nos. 11474113, 11004066 and 11274130, and by the Fundamental Research Funds for the Central Universities under Grant No. HUST: 2013QN014. We also acknowledge the helpful discussion with Prof. Ruqian Wu.

## Author contributions

G.D. and G.G. designed the outline of the manuscript and wrote the manuscript. G.G. supervised the whole work. G.D. carried out the numerical calculations. K.Y. contributed discussions. All authors reviewed the manuscript.

## Additional information

**Competing financial interests:** The authors declare no competing financial interests.

**How to cite this article:** Ding, G., Gao, G. & Yao, K. High-efficient thermoelectric materials: The case of orthorhombic IV-VI compounds. *Sci. Rep.* **5**, 9567; DOI:10.1038/srep09567 (2015).

# Author Queries

| Query Reference | Query |
|---|---|
| 1 | Equation (3) has been included twice in the text. Could you please check your equation numbering and ensure that each equation has a unique number? |